# Realization of an Economical Polymer Optical Fiber Demultiplexer

M. Haupt[1], C. Reinboth[2] and U. H. P. Fischer [1]

[1] *Harz University of Applied Studies and Research*
*Friedrichstraße 57-59, 38855 Wernigerode, Germany*
[2] *Innovations- und Gründungszentrum*
*Schlachthofstr. 4, 38855 Wernigerode, Germany*
*Email address: mhaupt@hs-harz.de*

*Abstract* - Polymer Optical Fiber (POF) can be and are being used in various fields of applications. Two of the main fields are the automotive and the home entertainment sector. The POF can be applied in several different optical communication systems as automotive multi-media busses or in-house Ethernet systems.

The requirements of bandwidth are increasing very fast in these sectors and therefore solutions that satisfy these demands are of high actuality. One solution is to use the wavelength division multiplexing (WDM) technique. Here, several different wavelengths can carry information over one POF fiber. All wavelengths that are transmitted over the fiber, must be separated at the receiver to regain and redirect the information channels. These separators are so-called Demultiplexers.

There are several systems available on the market, which are all afflicted with certain disadvantages. But all these solutions have one main disadvantage, they are all too expensive for most of the applications mentioned above. So the goal of this study is to develop an economical Demultiplexer for WDM transmission over POF.

The main idea is to separate the chromatic light in its monochromatic components with the help of a prism with low reciprocal dispersive power. The prism and the other assemblies, which are needed to adjust the optical path, should be manufactured in injection molding technique. This manufacturing technique is a very simple and economical way to produce a mass production applicable Demultiplexer for POF.

## I. INTRODUCTION

Polymer Optical Fibers (POF) have the power to displace and replace traditional communication systems via copper or even glass fiber in short distances.

One main application area is the automotive industry. There, POF displaces copper step by step because of its lower weight. Another reason is the nonexisting susceptibility to any kind of electromagnetic interference. These two advantages render optical communication systems first choice for the automotive industry.

Furthermore POF offers easy and economical processing and is more flexible for plug packing compared with glass fiber. POF can be passed with smaller radius of curvature and without any disruption because of its larger diameter in comparison to glass fiber.

Another sector where POF applies for communication is the multimedia in-house Ethernet system, as shown in fig. 1, [1], [2].

Here different application scenarios can be applied, which are mainly parted in three fields:

- "A/V Server Network" (communication between e.g. television, hi-fi-receiver and DVD-player)
- "Control Server Network" (messaging between e.g. refrigerator and stove)

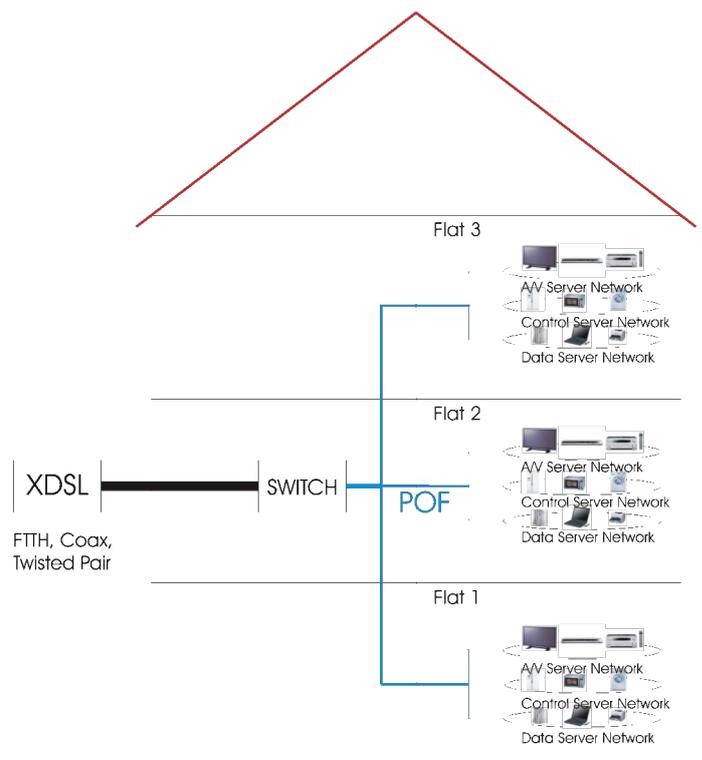

Fig. 1 Local Multimedia Infrastructure

- "Data Server Network" (data exchange between e.g. notebook and printer)

All these services and applications provide a large amount of data which must be carried for communication. Even communication via polymer optical fiber is limited by 2 Gbit/s.

Hence new ways of data transmission should be found to master these high bandwidth applications. One promising attempt is to use more than one wavelength to carry information via optical fiber. This technique is called Wavelength Division Multiplexing (WDM), [3], [4], [5]. There light consisting of various wavelengths is carried simultaneously over one single optical fiber. Every single monochromatic part of this propagating light carries information. Hence there is no limitation in bandwidth for optical fiber using WDM.

But two new parts must be integrated in the communication system. The first is the Multiplexer which must be placed before the fiber to integrate every wavelength to a single waveguide. The second component, the Demultiplexer, is placed after the fiber to regain every discrete wavelength. Therefore the polychromatic light must be splitted in its monochromatic parts to regain the information.

This technology has the power to master the bandwidth requirements which are needed to provide new multimedia applications in various fields of life.

## II. WDM Demultiplexer

Each commercial available WDM Demultiplexer performs after one of the following principles:
   a) Arrayed Waveguide Gratings (AWGs), this technology is only applicable for infrared range and multi-mode fibers.
   b) Fiber Bragg Gratings (FBGs) are only available for infrared range.
   c) Thin-Film Interference Filters are only available for infrared range as well.

The configuration of the new WDM Demultiplexer is shown in fig. 2, [6], [7], [8]. Light is carried via a standard step index polymer optical fiber (SI-POF) with a core diameter of 0,98mm and a cladding thickness of 0.01mm. Therefore the standard POF is 1mm in diameter. The core material consists of PMMA (polymethylmethacrylat) with a typical refractive index of $n_{PMMA}$=1.49 in the visible range. The cladding consists of fluorinated PMMA with a slightly lower refractive index. The numerical aperture shows values of 0.5 and hence the emitted light beam has a divergence angle of 30°. To separate the information carried by the monochromatic parts of the light, the divergent beam has to be separated and focussed. In this principal configuration a concave lens is applied to focus the light. The prism with low reciprocal dispersive power separates the several colors of light. The goal is to separate the different wavelengths on the "Detection Layer" in the size of a few millimetres. This separation should be adapted to an opto-electrical detector, which is situated in the point of focus to get the information without any cross-talk.

The sketch shows a basic setup with only three colors: red, green and blue. There is no limitation in reality, but for the first configuration it is useful to reduce the transferred wavelengths.

This principle configuration was simulated with the help of computer simulation software (OpTaliX).

One of the early results is shown in fig. 3. The refraction power to focus the light is divided by two lenses. The use of two lenses gets better results than the use of one lens due to aberrations.

A single biconvex lens shows many aberrations, e.g. spherical and chromatic aberrations. Hence it is more useful to split the refractive power by two lenses. The result is a lower spherical aberration, because of the lower radii which are needed with two lenses to achieve the same refractive power. The chromatic aberrations are reduced by using plano-convex lenses. A welcome side-effect is produced by the first lens: collimation of the light. A collimated light beam reduces the aberrations for a prism. This prism shows a different dispersive power for different wavelengths. The splitting is higher if the refractive index is of high value and differs strongly in comparison with the wavelength. In general at lower wavelengths higher refractive index are realized and vice versa. The more the gradient of the curve the better is the separation of every single wavelength. Fig. 4 shows different characteristics of four typical optical materials of refractive indexes in relation to the wavelength of the visible spectrum of light.

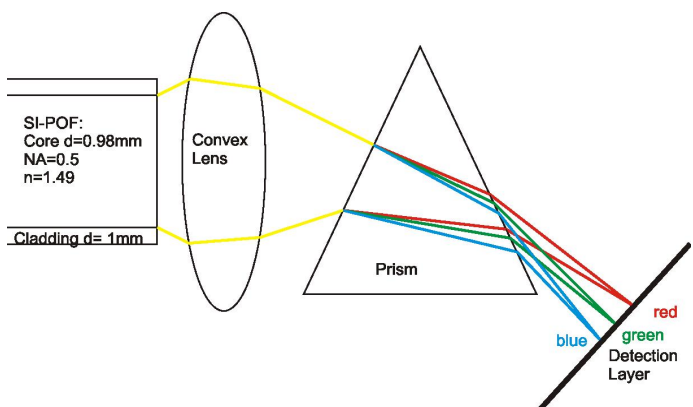

Fig. 2 Principal Sketch of a WDM Demultiplexer

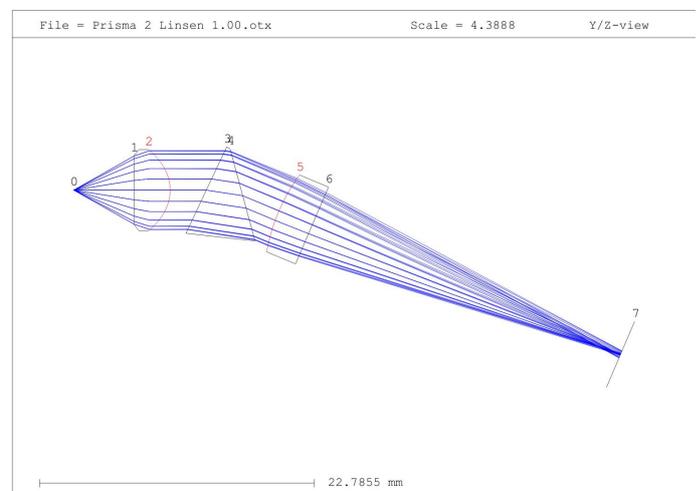

Fig. 3 2D Plot of early simulation

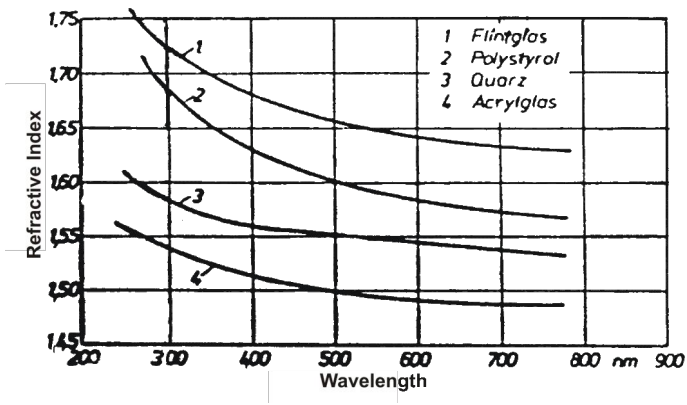

Fig. 4 Refractive index in dependence of wavelength

In fig. 3 layer "0" shows the emitted light of the POF. It can be considered as a point source, if the divergence angel and the diameter of the core are included. The first lens consists of the layers "1" and "2", it is a plano-convex lens to reduce aberrations. The low reciprocal dispersive power prism is situated between the two lenses and consists of the layers "3" and "4". The second plano-convex lens, layers "5" and "6", focuses the out of the prism escaping light on a detection layer "7". On the detection layer, there must be enough space between every single point of focus to detect the various wavelengths with the help of an electro-optical detector.

The first results shown are simulated at the very beginning of the analysis. To underline the result of this configuration a spot diagram for the detection layer is shown in fig. 5. A spot diagram collects the transverse aberrations in the image plane resulting from tracing a rectangular grid of rays (emerging from a single object point) through the system. As this analysis method shows, the different colors cannot be separated completely. Only two of the three colors can be separated. The red color with a wavelength of 660nm and the blue color with a wavelength of 470nm can be separated only with overlap and high cross-talk. The green color with a wavelength of 530nm shows the same behaviour.

The aberrations of the two lenses and the prism are too strong and there is no consistent point of focus. The reason of this behaviour is that the focus is shifted along the optical axis and therefore the diameter of the spot of every wavelength especially for the red and blue color is too large.

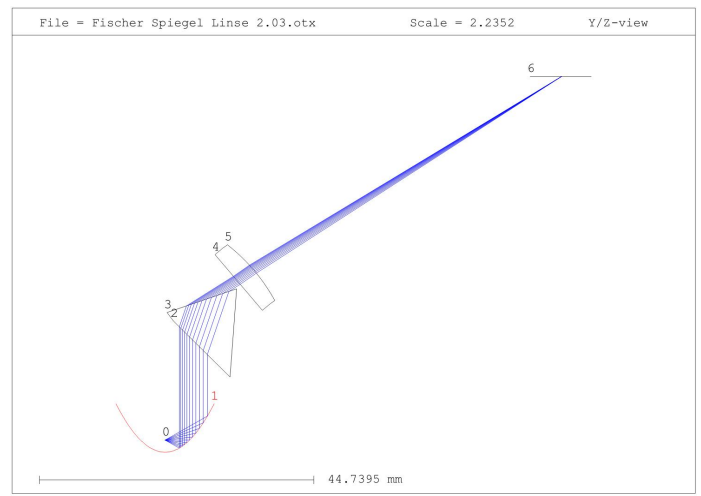

Fig. 6 2D Plot of improved simulation

A second configuration tries to reduce eminently the chromatic aberrations. The basic difference in the configuration is the use of a mirror instead of a lens to collimate the divergent light beam. The 2D Plot is shown in fig. 6. The mirror is a parabolic off-axis mirror. A parabolic mirror collimates the light to a perfect parallel light beam emitted by a light source which is situated in the focus point of the mirror.

A mirror has one main advantage compared to a lens; there is no chromatic aberration, because the light passes no other material with a different refractive index.

Hence the light caroming the prism is free of chromatic aberrations.

Again layer "0" is the source. The improvement and the change of configuration is layer "1", the off-axis parabolic mirror. This mirror is tilted by 90° and therefore the rays hit the concave mirror not on-axis in the angular point. The perfect collimated light is separated in its monochromatic parts with the help of the prism, layers "2" and "3". The only lens in this configuration, layers "4" and "5", focuses the rays onto the detection layer "6".

The base area of the whole configuration is smaller than $6\times10\,cm^2$. Hence it can be considered as a compact solution for a WDM POF Demultiplexer. This solution should produce better results. In comparison with the first simulation the spot diagram is shown in fig. 7 as well. One reason for the better result is the reduced chromatic aberration. The second is the path length of the rays through this configuration. As the distance between layers "5" and "6", the detection layer, is increased the gap between the single points of focus is also increased.

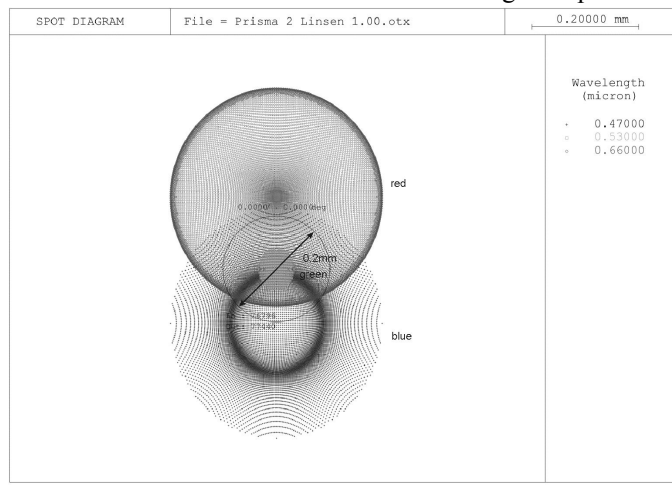

Fig. 5 Spot Diagram of Detection Layer

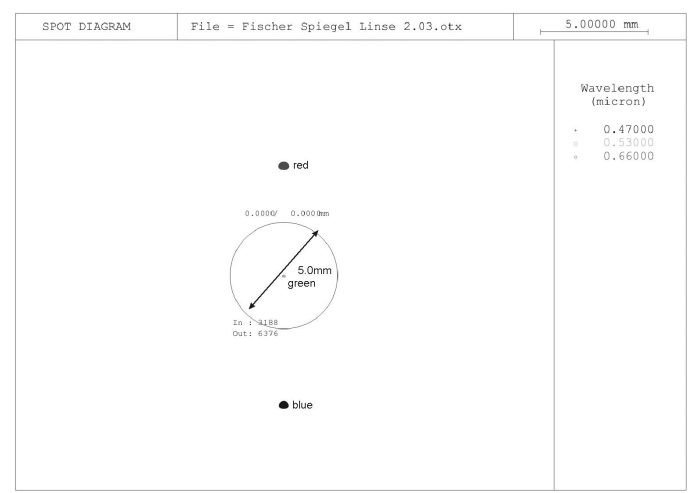

Fig. 7 Spot Diagram of Detection Layer of improved configuration

The result, as fig. 7 shows, is a gap of about 5mm between the red and the green color and the green and the blue color. This gap is large enough to detect and regain the information sent via the POF with the help of a photo-detector.

### III. RESULTS

The goal of the project is to develop a new economical way to increase the bandwidth of polymer optical fiber. This is necessary because of the increasing demand of high-speed communication systems for e.g. automotive or in-house applications. This demand can be satisfied with the help of wavelength division multiplexing, where it is possible to use more than one wavelength to carry information via an optical fiber. To apply this technique, it is essential to design an economical Demultiplexer. There are several systems available on the market, all with one main disadvantage; they are all too expensive for mass market.

Hence a new development is shown here. The main function of a Demultiplexer is the separation of the monochromatic parts of light. It is exploited that the refractive index of the used material is not a constant over the full spectral range, but rather depends on the wavelength of light, as it is shown in fig. 4. Therefore a prism with low reciprocal dispersive power can easily separate the different wavelength of light in different directions. This is the core idea for this Demultiplexer - to use a prism instead of wavelength selective mirrors or grids.

These presented results show, that it is a good way to design a Demultiplexer by means of a prism.

The first shown configuration has some improvements in comparison to the principal sketch. The refractive power to focus the divergent light beam emerging the POF is splitted into two plano-convex lenses to reduce spherical and chromatic aberrations. The occurrence of spherical and chromatic aberrations is so much the worse the stronger the radius of curvature of a convex lens.

To reduce these types of aberration, it is necessary to increase the radius of curvature, but this causes lower refraction power.

Hence to lower chromatic and spherical aberrations the refraction power is distributed in two plano-convex lenses. A comparison of spherical aberrations for different lens forms is shown in fig. 8.

As the results of the early configuration show, the aberrations are so strong, that the points of focus are shifted along the optical axis and therefore the spot-size of the different colors differs extremely in the detection layer.

The spot size is about 0,5mm in diameter for the blue and the red color. The low reciprocal dispersive power of the prism is too weak to separate the three colors. Hence only two colors can be regained. The green color in the middle is overlapped by the red and by the blue spot.

Therefore a new configuration must be designed to separate the three colors completely. Two basic attempts must be applied:
  a) The first is to reduce the chromatic and the spherical aberrations again.
  b) The second is to optimize the form of the prism to achieve better local separation of the applied wavelength.

These two goals are accomplished with the new configuration.

To reduce the spherical and chromatic aberration to a value of zero, an off-axis parabolic mirror is used instead of a lens.

A mirror has one main advantage, because the light is not passing another medium with a different refractive index, there cannot be any chromatic aberrations.

To avoid spherical aberration, the characteristic of a parabolic mirror is exploited. Light emerges the aspheric mirror in a perfect collimated beam, if the light source is placed in the point of focus of the mirror. Hence chromatic and spherical aberrations are non-existent.

The second idea to separate every single wavelength is to optimize the shape of the prism by using different values of angels.

If the light is diffracted stronger the gap on the detection layer between the single colors increases.

These steps increase the gap of every single part of light dramatically. The gap between the colors is about 5mm in length (see fig. 7). Hence they can be easily detected by opto-electronic detectors to process the transferred information. For that size of gap cross-talk is absolutely negligible (<< 30dB).

Another possibility to gain greater gaps is to optimize the material of the prism. If the refraction power is stronger and the gradient of the curve shown in fig. 4 is higher, the results are greater gaps as well. Hence in the second configuration the prism is made of PC (polycarbonate, $n_{PC}=1.59$). And the Abbe number is about 30. The Abbe number shows the power of dispersion of a material. The lower the Abbe number the higher is the dispersion and the gradient of the curve shown in fig. 4.

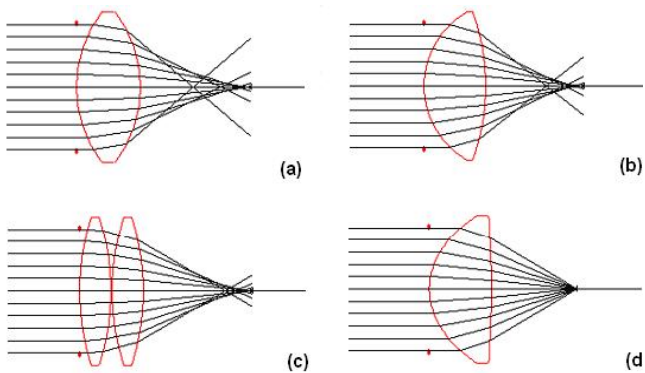

Fig. 8 Spherical Aberration for different lens forms: a) simple biconvex lens, b) lens "best form", c) distribution of refraction power in two lenses, d) aspheric, almost plano-convex lens [9]

The prism is applied of a plastic material as well as the fiber. In the second configuration - in contrast to the first - the lens, which focuses the light on the detection layer, is made of PMMA. Therefore every component of the configuration is made of a polymeric material. This is an enormous advantage because these components can be fabricated in a very simple and economical process: the injection molding technique. This manufacturing technology has the power to open this Demultiplexer for mass market and that is the goal.

As mentioned above, there are several demultiplexing systems available on the market, but they are all too expensive for most of the applications shown in the introduction. This configuration shown here in alliance with injection molding technique can create economical Demultiplexer.

## IV. CONCLUSION

There are many applications e.g. in the automotive sector or in the in-house communication which require communication systems with high data throughput. These demands grow almost daily. Hence new ways of data transferring methods must be found to satisfy all application demands. One auspicious way is to combine the easily manageable and processable POF technology with the economical injection moulding technique to use wavelength division multiplexing instead of only single wavelength technique via optical fiber. Single wavelength transmission over POF can achieve data rates up to 2Gbit/s. This limitation can be overcome by several wavelengths carrying information via the fiber. WDM requires Multiplexers and Demultiplexers. Demultiplexers can be designed with optical grids or mirrors to separate the different wavelengths again. These methods are very expensive and therefore not useable for most applications mentioned above.

This paper shows a Demultiplexer with a prism. The results show, that it is possible to design such a configuration. Even the early simulation shows results that satisfied the demand for a Demultiplexer, but these results have to be further developed before using them in any practical application. For that reason the second configuration has many advancements e.g. an aspheric mirror instead of a lens. These ameliorations show greater size of gap between every single wavelength in the detection layer. This causes easy detection for opto-electronic detectors.

These results alone are not enough to open WDM over POF for mass market. Only in combination with polymeric materials for the elements of the configuration and the fabrication in injection moulding technology, is it possible to achieve unit prices acceptable for the broad mass market.

In conclusion, WDM over POF is the solution for the increasing demand of bandwidth for all fields of applications.

An inexpensive Demultiplexer can be made by means of injection moulding technique and hence it is possible to use this Demultiplexer in many applications where high bandwidth is required.

The next steps to develop this demultiplexing technology ready to market are to manufacture a prototype to approve the simulated results.

## ACKNOWLEDGEMENT

We have to thank the State of Saxony-Anhalt and especially the State Secretary of Education for the "OPTOREF" project within the State Excellence Program.